\documentclass[10pt,letterpaper]{article}
\usepackage[top=0.85in,left=2.75in,footskip=0.75in,marginparwidth=2in]{geometry}

\usepackage[utf8]{inputenc}

\usepackage{cite}

\usepackage{nameref,hyperref}

\usepackage[right]{lineno}

\usepackage{microtype}
\DisableLigatures[f]{encoding = *, family = * }

\raggedright
\usepackage{longtable}

\usepackage{changepage}

\usepackage[aboveskip=1pt,labelfont=bf,labelsep=period,singlelinecheck=off]{caption}

\makeatletter
\renewcommand{\@biblabel}[1]{\quad#1.}
\makeatother

\usepackage{lastpage,fancyhdr,graphicx}
\usepackage{epstopdf}
\fancyheadoffset[L]{2.25in}
\fancyfootoffset[L]{2.25in}

\usepackage{color}

\definecolor{Gray}{gray}{.25}

\usepackage{graphicx}

\graphicspath{ {./figures/} }

\usepackage{sidecap}

\usepackage{wrapfig}
\usepackage[pscoord]{eso-pic}
\usepackage[fulladjust]{marginnote}
\reversemarginpar

\usepackage{amsmath}

\usepackage{gensymb}

\usepackage[binary-units = true]{siunitx}
\DeclareSIUnit\px{px}

\usepackage[toc,page]{appendix}

\begin{document}
\vspace*{0.35in}

\begin{flushleft}
{\Large
\textbf\newline{Reconstructing the visual perception of honey bees in complex 3-D worlds}
}
\newline
\\
Johannes Polster\textsuperscript{1},
Julian Petrasch\textsuperscript{1},
Randolf Menzel\textsuperscript{2},
Tim Landgraf\textsuperscript{1*}
\\
\bigskip
\bf{1} Dahlem Center of Machine Learning and Robotics, Institute for Computer Science, Freie Universität Berlin, Berlin, Germany
\\
\bf{2} Institute for Neurobiology, Freie Universität Berlin, Berlin, Germany
\\
\bigskip
* tim.landgraf@fu-berlin.de

\end{flushleft}

\section*{Abstract}
Over the last decades, honeybees have been a fascinating model to study insect navigation. While there is some controversy about the complexity of underlying neural correlates, the research of honeybee navigation makes progress through both the analysis of flight behavior and the synthesis of agent models. Since visual cues are believed to play a crucial role for the behavioral output of a navigating bee we have developed a realistic 3-dimensional virtual world, in which simulated agents can be tested, or in which the visual input of experimentally traced animals can be reconstructed. 
In this paper we present implementation details on how we reconstructed a large 3-dimensional world from aerial imagery of one of our field sites, how the distribution of ommatidia and their view geometry was modeled, and how the system samples from the scene to obtain realistic bee views. This system is made available as an open-source project to the community on \url{http://github.com/bioroboticslab/bee_view}.


\section*{Introduction}
Honey bees are extraordinary navigators. They orient themselves in an area of several square kilometers around their hives and they communicate spatial properties of remote resources via the waggle dance \cite{von1967dance}. In the last decade, harmonic radar was used to trace the flights of navigating bees \cite{riley1996tracking}. Recent results suggest that bees can robustly find their nest, even with an invalidated path integrator achieved by displacing the animal in a black box - or disturbed sun compass - induced by pausing the internal clock via anesthesia \cite{cheeseman2014way}. Honey bees have been shown to perform shortcut flights between known and dance-advertised sites over novel terrain \cite{menzel2011common}, a behavior that indicates that geometrical relationships between location are represented in or computed by yet unknown neural structures. Experimental evidence for different strategies, such as path integration and visual guidance using picture memories, have been provided \cite{collett2002memory,srinivasan2014going}. However, it is still unknown how those components are combined and at which level of abstraction the different components are available to a navigating bee \cite{cheung2014still,cruse2011no}. 

While this question may ultimately be answered through electro-physiological studies, the flight behavior of navigating bees may provide clues about the nature of visual information that is used for a navigational tasks (such as finding back to the colony). Experimental studies that analyzed flight trajectories so far only looked at rather basic features, such as velocities or angles, which were then compared between treatments.  

We have mapped a large area (\SI{2.73}{\km\squared} in size) with a quadrocopter and have created a virtual representation of our test site’s visual environment. We implemented the imaging geometry of the honey bee’s complex eyes and are able to reconstruct the visual input available to a flying bee given her position in the world. Previously recorded flight trajectories of bees can now be replayed in the virtual world and hypotheses regarding the information bees use for a given navigational task can be tested.

In this paper, we present our implementation of reconstructing the bee’s view in the virtual world. We provide a detailed description of how our system performs with respect to runtime and imaging accuracy. We provide code and map data along with this paper. The software is available online on GitHub.

\section*{Previous Work}
Several models that mimic insect vision have been proposed with varying degree of realism. The compound eyes of insects are made up of thousands (in the case of the honeybee worker
about 5500) of hexagonally shaped ommatidia facing in different directions \cite{seidl_visual_1981}. Each ommatidium acts like a single eye with its own corneal lens and (in reference to the apposition eye) photoreceptor. But unlike the human eye, each ommatidium receives light from a very limited portion of the environment. An ommatidium thus can be thought of as one picture element, or pixel \cite{borst_drosophilas_2009}. 
\begin{figure}
\caption{Close up photographs of a bee's head \textit{Apis mellifera}. The individual hexagonally shaped facets and the central ocellus on top of the head can be distinguished. The compound eyes have an ellipsoid shape and are roughly parallel to each other. The strong curvature of the eye results in a large field of view (FOV): the honeybee has close to a 360$^\circ$ FOV, only limited by the region blocked by the thorax of the bee.}
\includegraphics[width=0.6\textwidth]{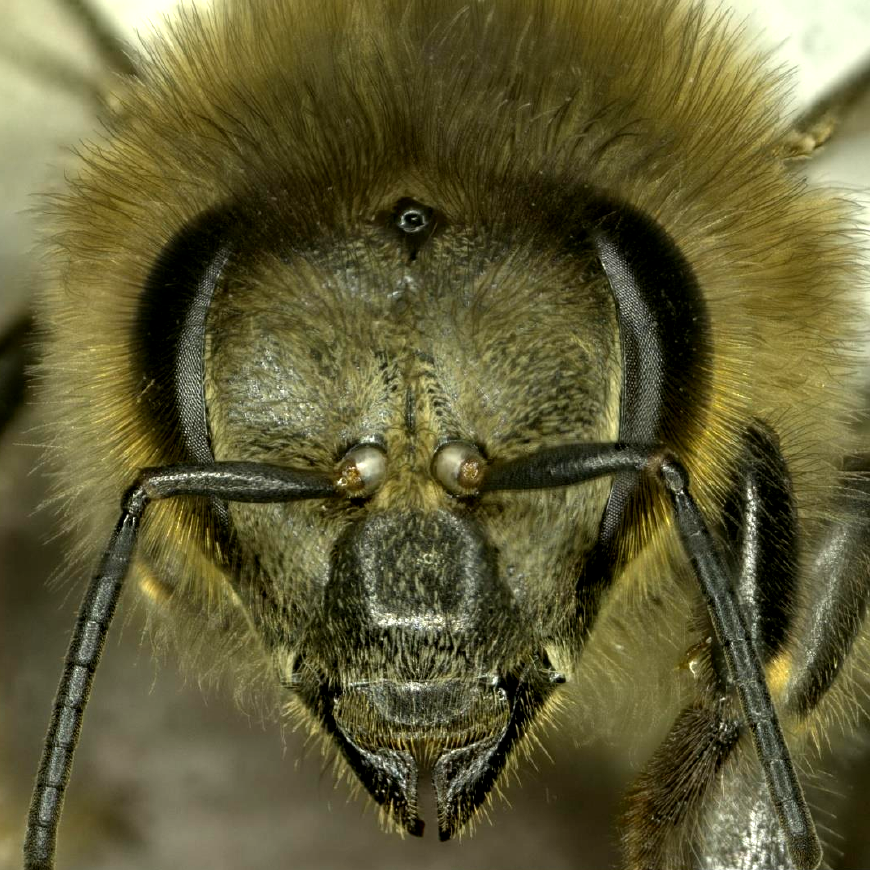}
\centering
\label{fig:beehead}
\end{figure}

\begin{figure}
\caption{Left: Interommatidial angles: The interommatidial angle ($\Delta \varphi$) is the angle between neighbouring ommatidia. One can differentiate horizontal (elevation, $\Delta \varphi_h$) and vertical (azimuth,  $\Delta \varphi_v$) angles. Right: The acceptance angle ($\Delta \rho$) defines the visual field of the individual ommatidia. The acceptance function describes the sensitivity of the ommatidium, in relation to the angular distance from the optical axis. The angular sensitivity function can be approximated by a two-dimensional circular Gaussian. The full width at half maximum (FWHM) of this Gaussian is the acceptance angle of the ommatidium  \cite{varela_optics_1970}.
}
\includegraphics[width=\textwidth]{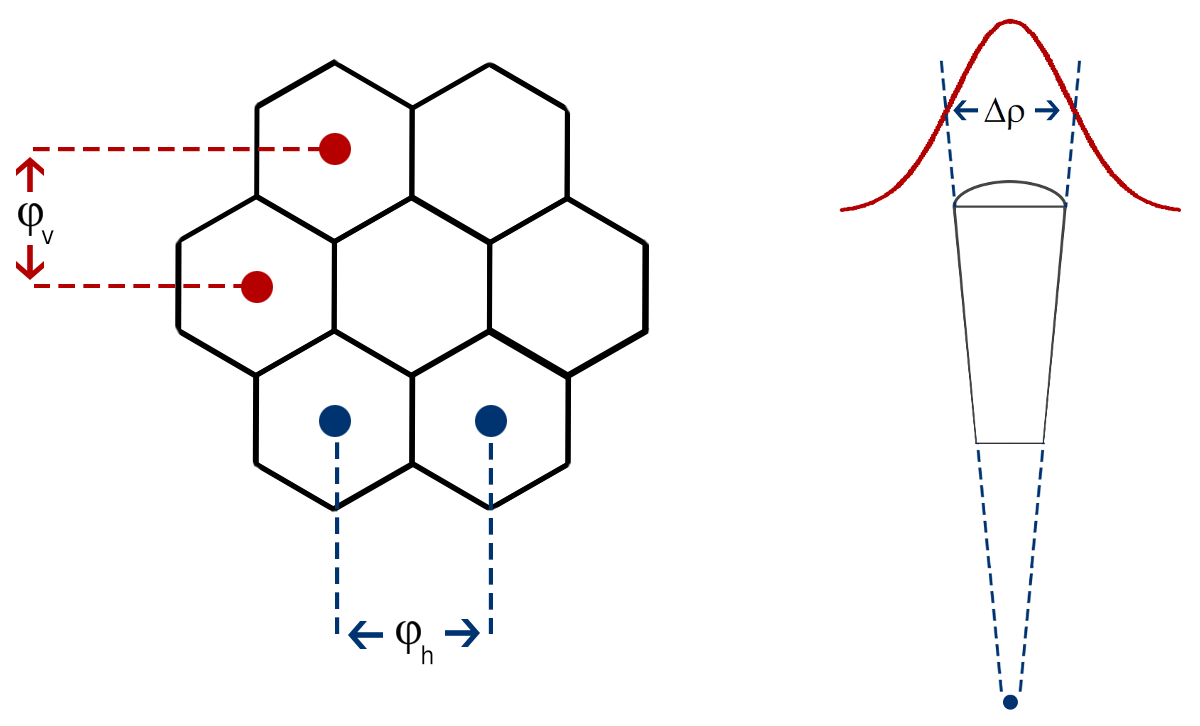}
\centering
\label{fig:acceptance_angle}
\end{figure}

Most relevant for a realistic imitation of the visual input are the viewing directions and the field of view for each ommatidium in the compound eye. Although the diameters of the ommatidia vary significantly, and therewith the receptor density over the compound eye surface, almost all models disregard this property, since the bee eye is comparatively small and the spatial resolution of the imaging process can be described reasonably accurately with two parameters  (see Figure \ref{fig:acceptance_angle}): interommatidial angles, which are the angles between the main viewing axes of neighboring ommatidia, and the acceptance angles, which reflects the field of view for each ommatidium. Both properties have been determined experimentally \cite{laughlin_angular_1971, seidl_visual_1981}. 
In table \ref{table:overview_models} table we have compiled a list of previously described insect vision models along with some of their properties. While each model exhibits it's own design decisions and particularities, a few basic distinctions can be made. While insect vision models as well might refer to hardware designs (like e.g. in \cite{floreano_miniature_2013}), we limited the list to mathematical models in virtual worlds. We can discriminate between works that use rather simple scenes such as 2-dimensional image planes or 3-dimensional geometric primitives. Recent works propose using 3-D reconstruction techniques such as photogrammetry or laser scanners to reconstruct realistic scenes (see e.g. \cite{sturzl_three-dimensional_2015}). Every model comes with a list of functionalities specific to the focal application, ranging from interfaces to neural simulators, physics engines to simulate environmental forces, or configurable spectral sensitivity. 

In our studies on honeybee navigation we have two target applications. First, we would like to analyze flight trajectories with respect to the animal's visual input \cite{Menzel_frontiers_2018}. Secondly, in a current project we record extra-cellular neural activity from honeybees on a quadrocopter. To investigate how the spike trains correlate with the animal's visual input, we require a realistic reconstruction of bee's perception. While substantial previous work has been done to reproduce the visual perception of bees, unfortunately, none of these solutions was either publicly available or able to use complex 3-D maps. Our goal thus was to implement an accurate model of the honeybee compound eye that can be executed in real-time, with an explicit raytracing instead of texture remapping. This model should be generically applicable to different eye models and different world models, and it should be freely available to the community. 

\section*{Implementation}
\subsection*{3-D World}
\begin{figure}
\caption{Illustration of how the 3D model is extended. A: Wireframe of the three Components. B: The cylinder with the panorama projected onto it. C: The hemisphere with the sky texture. Since the sky in this model is static it may introduce bias to the subsequent image analysis. We therefore also created a 3D model with solid white sky. D: The 3D Terrain captured by the drone. E: The resulting 3D model with skydome and panorama. }
\includegraphics[width=\textwidth]{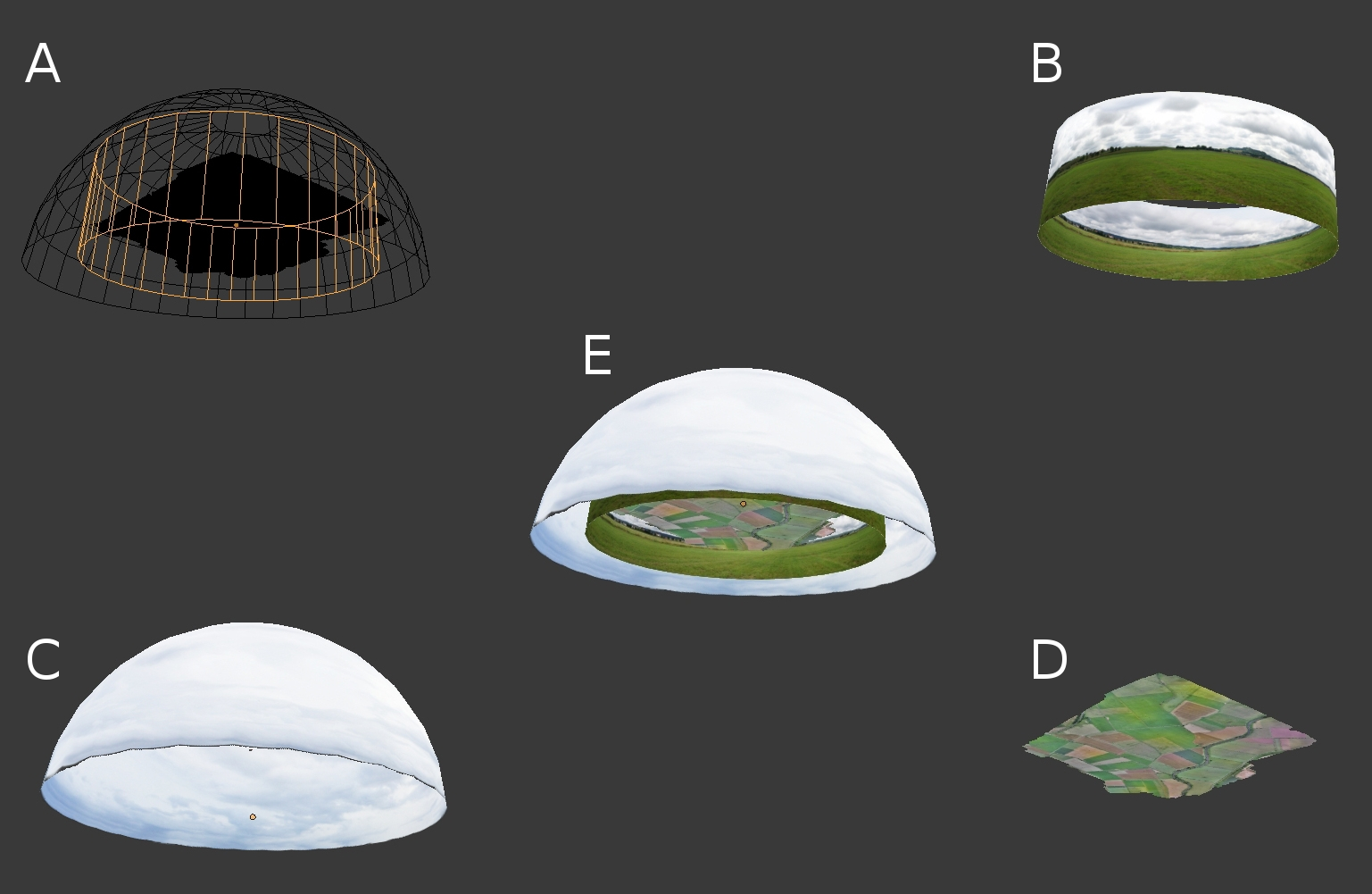}
\centering
\label{fig:collage_3D_model}
\end{figure}

The 3-D world consists of three parts: 1) a 3-dimensional depth map of an experimental field site surrounded by 2) a cylinder that holds a panorama image and 3) a sky dome (see figure \ref{fig:collage_3D_model}). The virtual world reproduced an area east of Großseelheim, a town in central Germany. It covers an area of about two square kilometers and was used for behavior experiments with bees over the last few years. The depth map was created in June 2016 from aerial images taken by a drone, using stereophotogrammetry. The resulting model has a vertical accuracy of 30 cm and a horizontal accuracy of \SIrange{5}{10}{\cm}. Therefore small bushes and trees appear with their respective shapes in the depth map but smaller objects such as fences and small plants are only visible in the texture of the model. The environment was highly structured and exhibits panoramic features that were too far away to be depth mapped by our drone. Hence, the model was extended by mapping a high resolution panoramic image onto a cylinder.

\begin{figure}[h]
\includegraphics[width=\textwidth]{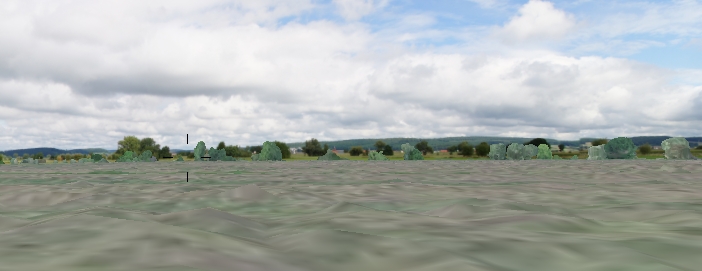}
\caption{Some objects in the 3D model also appear in the panorama map, e.g. the bushes in the background. Agents that are positioned far away from the location from which the panorama was recorded will perceive false object duplicates in the panorama.}
\centering
\label{fig:double_objects}
\end{figure}

Objects within the drone-captured area appear in this panorama texture irrespective of the camera position and a duplicate would be imaged to the bee eye, one from the actual 3-D object and one from the panorama texture. To solve this problem, duplicated objects were identified in the 3-D world and removed manually from the panorama with an image editing program. Larger objects (such as trees) were replaced with parts of other panoramas, since one can not see what is behind these objects (see Figure \ref{fig:double_objects}). Note that only one recording was used as panorama texture. It’s projection to the bee eye is correct only for positions close to the position at which we recorded the panorama. For all other positions in the world the projection exhibits an error proportional both to the distance of the original object’s position to the camera and the distance of the camera to the original panorama recording position. When moving closer to objects, the objects' projection grows larger, when moving away they appear smaller, when moving parallel to the objects they shift. In a \SI{360}{\degree} panorama such as our virtual world, all of these effects can be observed in any one move. However, since the area mapped by our drone is fairly large, the errors introduced by having only a static panorama are negligible. 
The resulting model has \SI{1000294} faces and \SI{499116} vertices. The resolution of the texture of the 3D terrain is $\SI{8000}{\px} \times \SI{8000}{\px}$, the resolution of the texture of the cylinder is $\SI{24000}{\px} \times \SI{3000}{\px}$. The model needs $\SI{472}{\mebi\byte}$ of disk space.

\subsection*{Raycasting}
In order to generate a realistic projection of the world’s object to our model of a bee eye, we cast rays from each ommatidium into the world. While “ray tracing” methods follow rays of light over multiple bounces off of scene objects, “ray casting” only takes into account the primary ray, i.e. only the light rays between camera and object are simulated. To achieve this, rays are generated from the camera. For each pixel of the image to be rendered, ray directions are calculated from the eye model. The rays are “shot” in the calculated directions. Then, every object in the scene is tested whether it intersects with the ray. This is computationally expensive since there can be millions of objects in a scene. After an intersection is found, the colour for the pixel is sampled from the object’s texture at the intersection point.

\subsection*{Model of the Bee's Compound Eye}
In the honeybee eye, the interommatidial angles vary across the bee’s compound eye, with a minimum at the equator (elevation = 0) 
and gradual increments towards the borders of the eye. This means, ommatidia have a smaller spacing, i.e. a higher resolution at the equator. Vertically, inter-ommatidial angles range from \ang{1.5} to \ang{4.5} and horizontally they range from \ang{2.4} to \ang{4.6}. For calculating the interommatidial angles, a routine described by Stürzl et al \cite{sturzl_mimicking_2010} was implemented. The routine is based on a formula from Giger \cite{giger_honeybee_1996}. Giger approximates the measurements of interommatidial angles determined by Seidl \cite{seidl_visual_1981} for all ommatidia in the frontal hemisphere. Stürzl et al. \cite{sturzl_mimicking_2010} extend this model to cover the full bee's eye FOV. They also take into account the border of the visual field. Since the authors of \cite{sturzl_mimicking_2010} did not provide source code, we re-implemented the routine as an R script\footnote{Also available on github: \url{https://github.com/BioroboticsLab/bee_view/blob/master/data/calc_ommatidial_array.R}}. 
This model produces angles for a total of \SI{5440} ommatidia per eye. These are precomputed and stored in a comma separated file for later use by the renderer. This way, the subsequent parts of the rendering pipeline can be used for updated models of ommatidium distribution or different animal models. 
These angles in 3-D space define the direction of the rays to be cast. Since individual ommatidia do not just register light coming in from this exact direction, but rather collect light from a field around this average vector, we need to define how much of the scene can be sampled by one ommatidium and with how much weight samples from differing directions are integrated into the output of an ommatidium. In \cite{sturzl_mimicking_2010}, the authors choose to use an acceptance angle that varies depending on the elevation and azimuth of the ommatidia. Since the interommatidial angles also vary, a static acceptance angle may lead to oversampling in areas of high resolution (e.g. the center of the eye) and undersampling (at the edge of the eye). The lens diameter also varies between \SI{17}{\micro\metre} and \SI{24}{\micro\metre} over the surface area of the eye, in \cite{sturzl_mimicking_2010} this has been interpreted as an indication for a dynamic acceptance angle, however as of now, there aren’t any direct electro-physiological measurements available for the whole eye. The only direct measurements were conducted in the frontal region of the eye and came up with an acceptance angle of \ang{2.6} \cite{laughlin_angular_1971}. In the model proposed by Stürzl and coworkers, the acceptance angles depend on horizontal and vertical interommatidial angles and hence are not radially symmetric. The model of Giger \cite{giger_honeybee_1996} implements a static acceptance angle of \ang{2.6} with a radially symmetric acceptance function -- an approach followed in our rendering engine.

\begin{figure}[h]
\caption{Gaussian sampling function with 462 samples showing the angular deviation from the main optical axis, the weights are shown as a third dimension.}
\includegraphics[width=\textwidth]{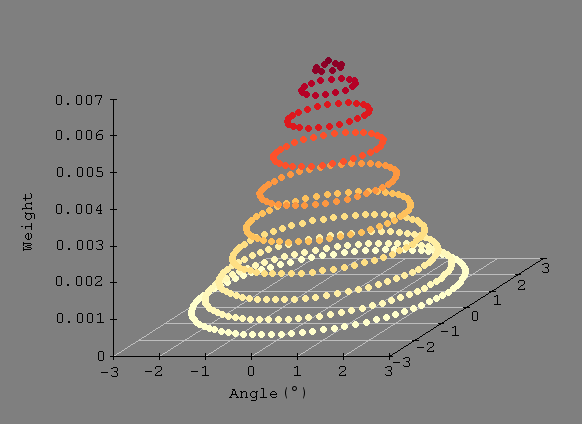}
\centering
\label{fig:3d_acceptance_function}
\end{figure}

\subsection*{Sampling from the Scene}
The acceptance function is a radially symmetric Gaussian with a full width at half maximum (FWHM) that is equal to the acceptance angle.
The Stürzl model uses $9\times 9$ sampling directions per ommatidium and weights the samples with a Gaussian weight matrix, whereas Giger uses a sampling array of 441 sampling points that are arranged as concentric circles around the optical axis of the ommatidium. Each sample is weighted according to its distance from the optical axis using a gaussian pdf with zero mean and a standard deviation of 1,1523 (in case of an acceptance angle of \ang{2.6}).

Similarly to Giger, we implemented a concentric disk sampling method. This is achieved by creating a square sample matrix with coordinates ranging from -1 to 1 and then mapping the sample points to a disk. Afterwards, the coordinates are normalized to be in the range of $-\Delta \rho$ to $+\Delta \rho$. The formula maps the x, y coordinates of a point in a square to the X, Y coordinates of a point in a disk \cite{rosca_new_2010}:

\[(x,y) \mapsto (X,Y) = \begin{cases} \left(\frac{2y}{\sqrt{\pi}}\sin\frac{x\pi}{4y} , \frac{2y}{\sqrt{\pi}}\cos\frac{x\pi}{yx}\right ), & \text{if } |x| \leq |y|\\\\ \left(\frac{2x}{\sqrt{\pi}}\cos\frac{y\pi}{4x} , \frac{2x}{\sqrt{\pi}}\sin\frac{y\pi}{4x}\right ), & \text{otherwise} \end{cases}\]

The acceptance function that weighs the sample points depending on the distance to the main optical axis of the ommatidium is given by: 

\[f(x,y) = e^{-\left ( \frac{5\sqrt{x^2+y^2}}{3\Delta p} \right ) ^2}\]

The formula approximates a bivariate Gaussian function with FWHM $\Delta \rho$. For a $\Delta \rho$ of \SI{2.6}{\degree} this equates to:

\[f(x,y) = e^{-0.410914\left ( x^2+y^2 \right )}\]

The weights produced from the formula are then normalized to sum up to 1. Figure \ref{fig:3d_acceptance_function} shows the acceptance function and corresponding weights for $N = 462$ samples. Figure \ref{fig:acceptance_angle_comparison} shows the acceptance function for other $N$, and also compares the differences between the sample points arranged in concentric disks, and the uniform square sampling method. 

\subsection*{Core Technologies}
The core renderer was written in C++. It uses Embree for intersecting the rays with the scene and the Eigen C++ Vector Library \cite{eigen_community_eigen_nodate} for fast vector arithmetic. Embree is a raytracing kernel developed by Intel and offers core raytracing functionality such as intersecting rays with the scene, while hiding the underlying acceleration structures and CPU optimizations. Additionally it has a good documentation and, even though still under development, the API is stable. Furthermore it is highly optimized for CPUs, achieving good results in benchmarks. Embree is free and open source, as it is released under the Apache 2.0 license. It runs on all modern x86 and AMD64 CPUs \cite{afra_embree_2016}. 

The rendering engine uses the Wavefront Object (.obj) file format, since it is supported by most of the major 3D applications and it is an open, human readable format.
The core renderer was wrapped in Python, as Python is widely used in scientific programming, so this provides an interface that can easily be used with other scientific applications. 
For every C++ function that should be wrapped, a corresponding Cython function was written that calls the C++ function. The result is the beeview python package that, after being built with the Cython compiler, can easily be imported to Python.

\begin{figure}[h]
\caption{Left image: an example plot of a flight from the release site     (RS) to the hive (H). Red is search flight, blue is linear flight and the dotted yellow lines are gaps. The yellow hexagon is the position of the bee. The central image shows the 2D bee view from that position (a $\SI{25}{\m} \times \SI{25}{\m}$ portion of the map), rendered with the Radar Track module. The right image shows the bee view rendered with the help of the beeview python package.}
\includegraphics[width=\textwidth]{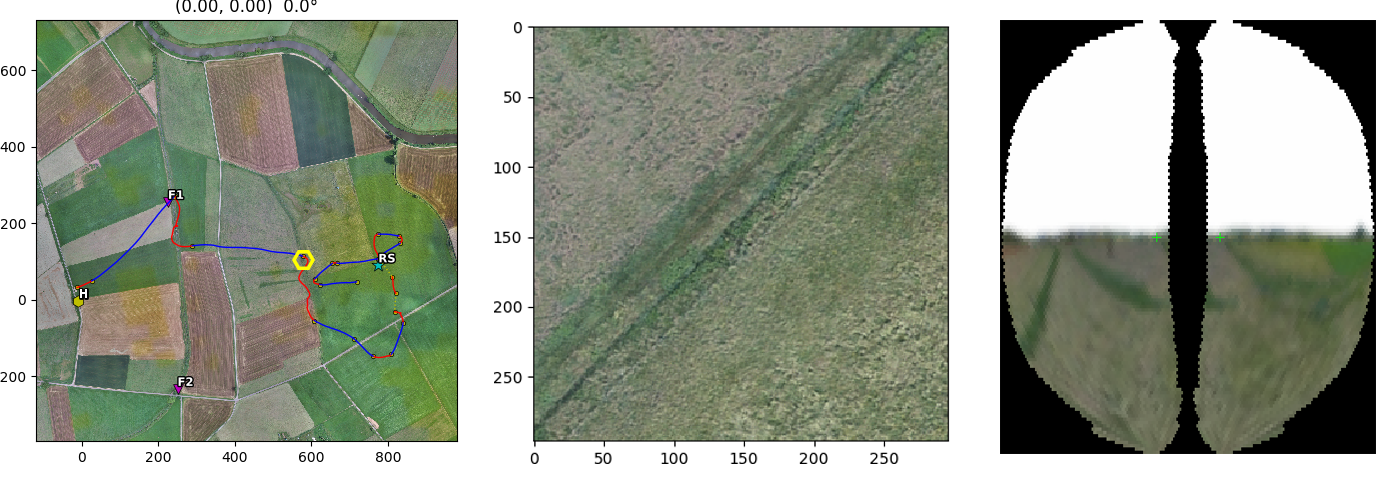}
\centering
\label{fig:radartrack_module}
\end{figure}

The  source code is available on Github\footnote{\url{https://github.com/BioroboticsLab/bee_view}}.

\section*{Results}
\begin{figure}[h]
\caption{Measurements for how long it takes to render one beeview for different sample sizes (per ommatidium) on different machines (a laptop with 2012 dual-core i7-3520M and a desktop with 2013 quad-core i5-4430). We rendered images from within the 3D Model (\num{e6} faces) and from a simple test scene (cube with 6 faces). Rendering on the quad-core CPU is roughly twice as fast as on the dual-core CPU and the render duration increases linearly with the number of samples. Also the number of faces of the scene has an impact on the rendering speed. For 56 samples in the large 3D world it takes 260ms to render a beeview.}
\includegraphics[width=\textwidth]{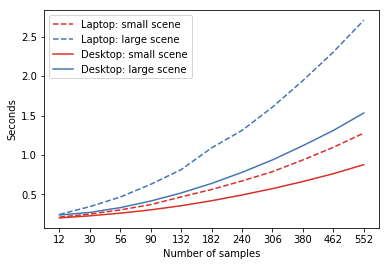}
\centering
\label{fig:benchmark}
\end{figure}

In this section we look at how different rendering parameters affect the output and the performance of the renderer.
The runtime performance of raycasting directly depends on the number of rays to cast and the amount of polygons in the scene. These are given by the bee’s eye model and the scene described in the section “3D World”. The number of rays needed for rendering a bee view is $2N_oN_s$. Where $N_o$ is the number of ommatidia and $N_s$  is the number of samples per ommatidium. For 462 samples per ommatidium the renderer generates \SI{5026560} rays and performs as many intersection tests. The scene has over \num{e6} faces that need to be tested for intersection. See Figure \ref{fig:benchmark} for a benchmark on how these parameters effect render speed.

\begin{figure}[h]
  \centering
\caption{These renderings show a portion of the bee view (the red rectangle) for different sample sizes to determine how the number of sample points affects the simulated bee vision. For this a series of bee views was rendered, with the number of sample points per ommatidium ranging from 12 to 9900. Until n = 56 the differences between the bee views are quite large. From n = 56 to n = 462 the image still becomes smoother, but the differences are almost not noticeable. From n = 462 there is no conceivable difference between the rendered bee views.
}
\includegraphics[width=0.75\textwidth]{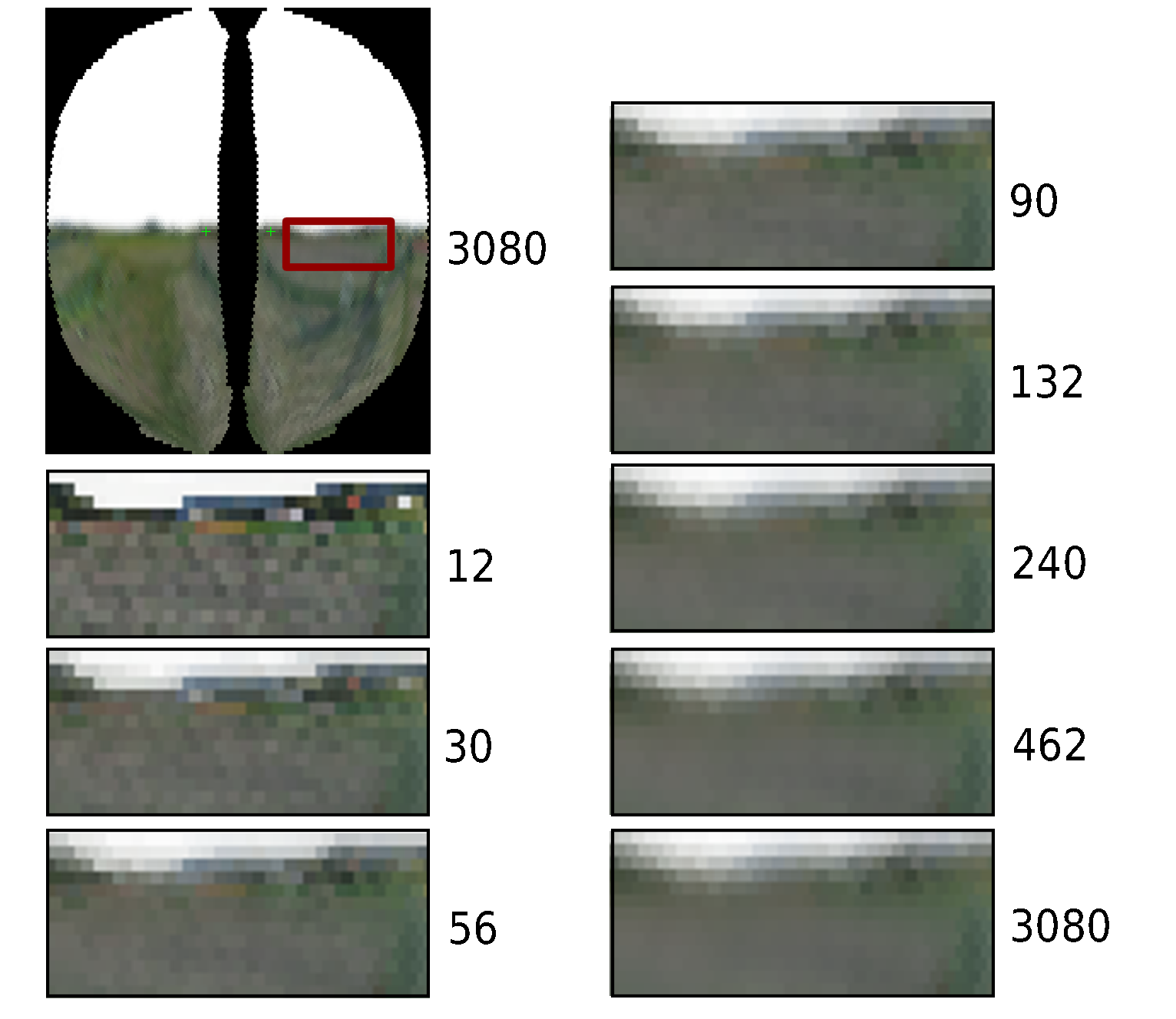}
\label{fig:beeviews_ausschnitt}
\end{figure}

Since the performance of the renderer directly depends on the number of rays to cast, users might decide decreasing the number of rays for faster rendering. We conducted a test series to determine the minimal sample size per ommatidium at which the renderer still yields acceptable results (see Figure \ref{fig:beeviews_ausschnitt}). From visual inspection we conclude that more than 56 rays per ommatidium may not be necessary. Lower numbers decrease rendering times but produce choppier images. 

\begin{figure}[h]
\caption{The effect of different acceptance angles on the output of the renderer. left: \ang{1.3}, centre: \ang{2.6}, right: \ang{5.2}. A smaller acceptance angle leads to a sharper image.}
\includegraphics[width=\textwidth]{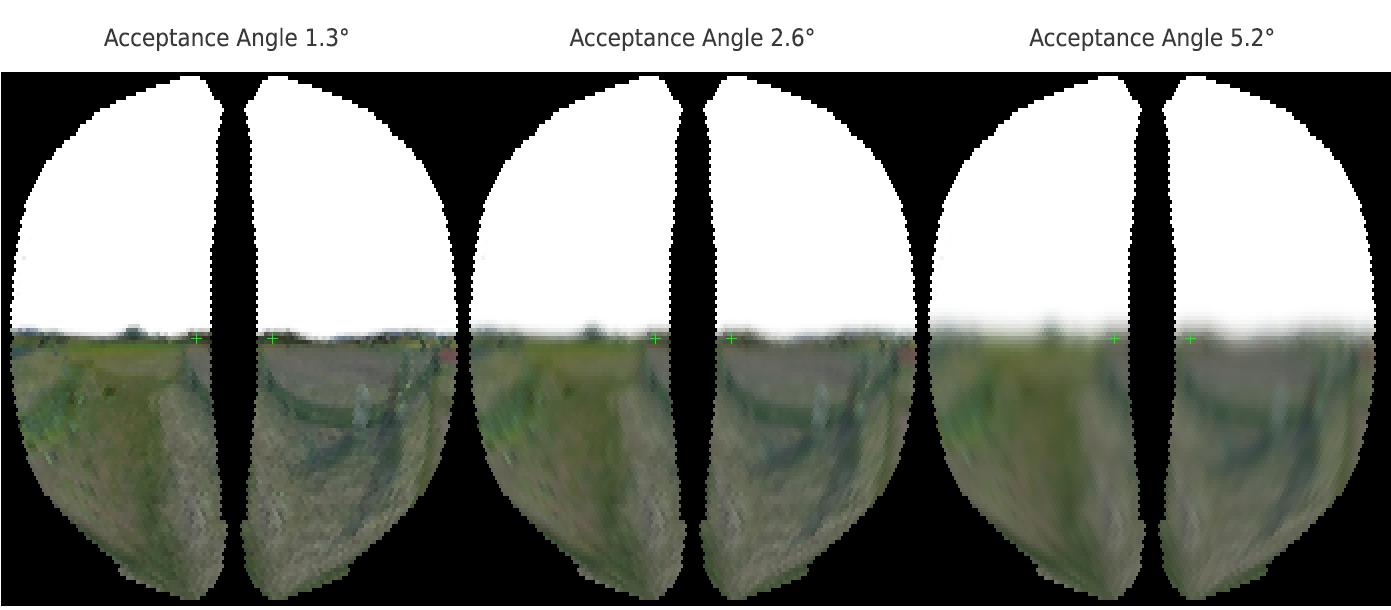}
\centering
\label{fig:acceptance_angle_comparison}
\end{figure}

The acceptance angle controls the sharpness of the rendered beeview (see Figure \ref{fig:acceptance_angle_comparison}). A larger acceptance angle leads to a blurrier Image. Also, objects that are farther away are not as sharp as closer objects, since with greater distance the acceptance angle covers a larger area.

\begin{figure}[h]
\caption{Comparison of differences between using a square sampling     distribution.}
\includegraphics[width=\textwidth]{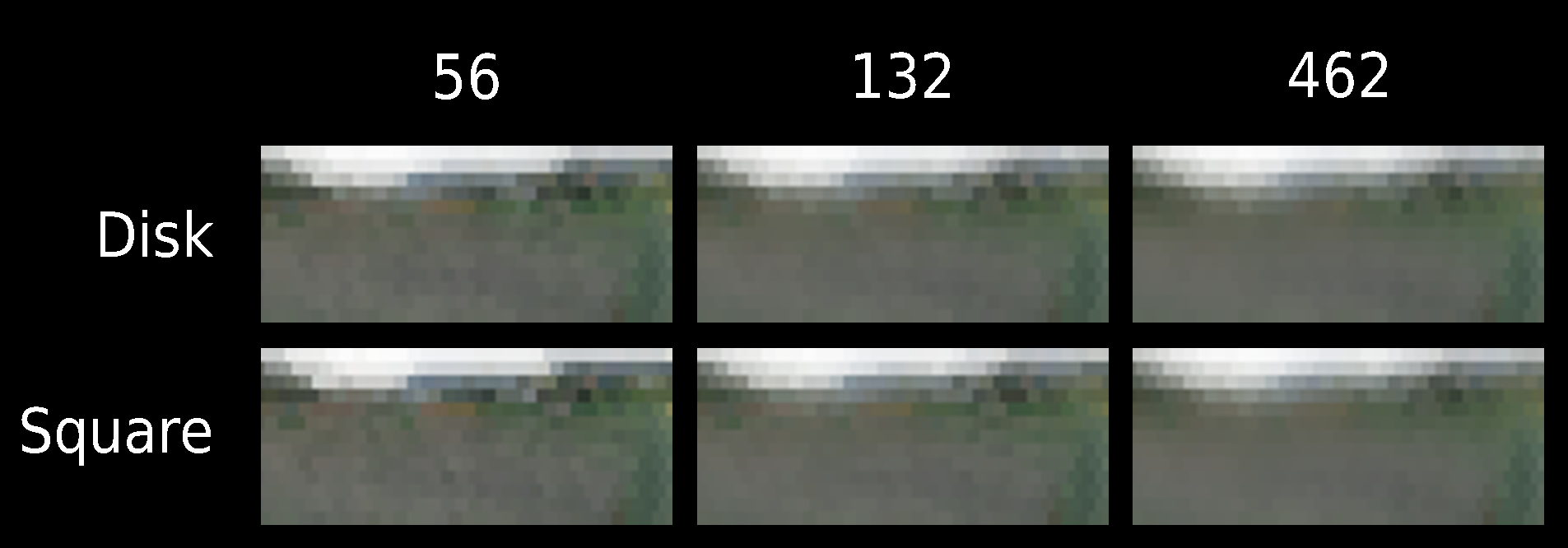}
\centering
\label{fig:comparison_quadratic_concentric}
\end{figure}

We also explored if and how the rendering output is affected by using a square sampling distribution (similar to the Stürzl-model \cite{sturzl_mimicking_2010} or a concentric disk distribution similar to Giger \cite{giger_honeybee_1996} (see Figure \ref{fig:comparison_quadratic_concentric}. We find that there is no perceivable difference between the two methods, except for small sample sizes, as square sampling covers a larger area (since the  width of the square and the diameter of the disk are equal to $2 \Delta \rho$). However for larger sample sizes the differences are less pronounced, since the samples at the edge contribute with small weights.

\begin{figure}[h]
\caption{These renderings show the simulation of how a bee sees a $\SI{2}{\meter} \times \SI{2}{\meter}$ image from different distances. Image A: The test scene, rendered with the pinhole camera. The test image shows squares of different colours with numbers in them. Each square has a width and height of \SI{25}{\cm}. Because of the square form, distortions are easily identified. The numbers and colours help distinguish the different squares. As mentioned before, the renderer only reproduces the sampling geometry and not the spectral perception of honey bees. The colors, hence, serve purely for distinguishing the squares. Image B: The test image seen from a distance of \SI{2}{\meter}. Image C: \SI{1}{\meter}. D: \SI{50}{\cm}. E: \SI{25}{\cm}. F: \SI{10}{\cm}. The settings for the bee camera used: acceptance angle  \SI{2.6}{\degree}, disk sampling, 132 samples per ommatidium.}
\includegraphics[width=\textwidth]{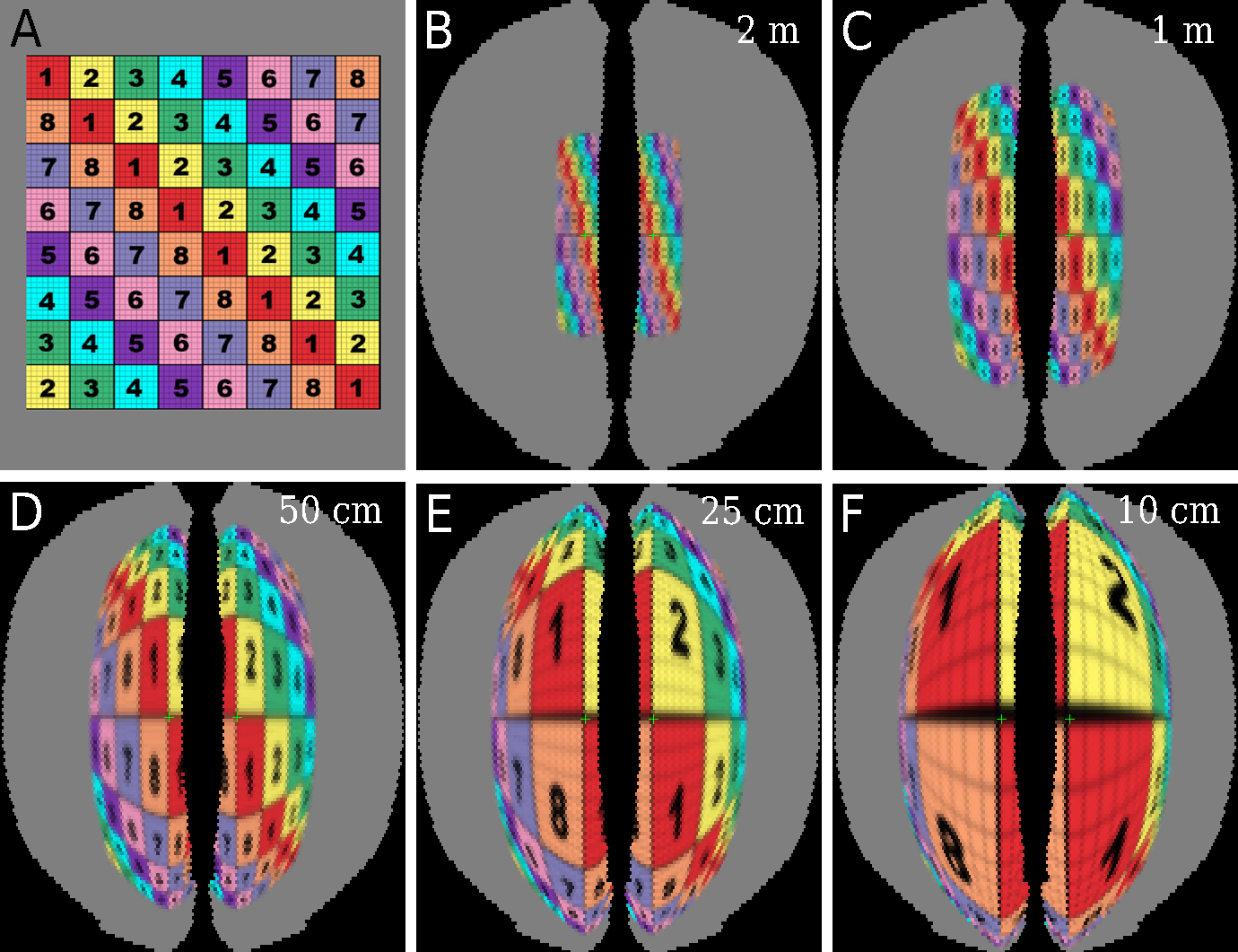}
\centering
\label{fig:beeview_test_series}
\end{figure}

We conducted a series of tests to examine the properties of the rendered bee views (see Figure \ref{fig:beeview_test_series}). The test series shows that objects in the centre of the eye appear enlarged, since the resolution of the eye is highest here. Also, the closer the object, the more it appears distorted, because the object covers a larger part of the field of view. Additionally the model confirms that only a small portion of the field of view of the eyes overlap.
 
\section*{Discussion}
We have implemented a fast, accurate and open software package to reproduce the visual perception of honeybees. It is the first system of this kind made available as open-source package.

The renderer is not limited to rendering bee views, but can also render normal images from a pinhole or a panoramic camera. It has a simple API so it can easily be interfaced with from other applications. The beeview Python package provides bindings to the C++ functions of the renderer. The renderer is implemented in a way that all the settings of the renderer are flexible. This means that the renderer can easily be used with different interommatidial and acceptance angles for simulating the vision of other insects. 
The optimal settings for rendering bee views were determined so the performance is maximized while still maintaining accuracy.
With the proposed system, behavioral studies that investigate the relation of visual input to behavioral output can benefit from this system. The 3D model is also an ideal environment for synthetic studies using artificial agents. The C++ API and the beeview Python package provide all the functions needed for the movements of an agent like the one implemented by Müller \cite{muller_familiarity_2016}. Functions for moving, rotating and rolling the camera, so all the possible movements of a bee are covered. Furthermore it is possible to set the camera direction directly, or via a look\_at(point) function.
A render function that returns the elevation angles, the azimuth angles and the sampled colours of all ommatidia as continuous arrays for visual input of the agent. A function for measuring the distance from a point to the next object in a specific direction, that can be used for measuring the height above ground and setting the camera’s position accordingly.
The renderer can also be used for evaluating pattern and shape recognition experiments, as in \cite{giger_honeybee_1996}, \cite{srinivasan_pattern_1994}, \cite{anderson_shape_1977}, by placing test images at a specific distance from the camera.
Additionally the renderer can be used for educational purposes to demonstrate how different parameters of the compound eye affect insect vision.

Still, a number of aspects can be improved.
Loading the high resolution textures of the model has the biggest impact on the start-up time of the renderer (about \SI{10}{\s}). Using a faster image library instead of the simple ppm loader could speed up the process.  Only supporting the ppm file format for images is not optimal, but can easily be improved by using a different image library.
The model eye only takes into account the spatial resolution of honeybee’s eyes. The model could be extended to include the light intensity received by the ommatidia and the spectral sensitivity of the ommatidia. To achieve this, the 3D environment has to include UV emission information. This could be done by recording the scene with a camera sensitive to UV-light and storing the recorded UV-data in the red channel or alpha channel of the texture. Additionally only the compound eyes were modelled, but for a complete bee vision simulation the ocelli should also be taken into account.
The 3D model could be extended by including light sources and material properties. Based on these the renderer could render shadows and other light effects by tracing the rays for multiple bounces. The polarization of light could be simulated, as some ommatidia of the bee’s eye are sensitive to it.
The scene could be refined by using a subdivision mesh (Embree is capable of handling subdivision meshes) \cite{benthin_efficient_2015}. 
In the summer of 2017, we recorded a larger 3-D model. Embedding it in a DEM would be a good method for expanding the model, since the new model covers all scene objects that are near the testing area, and the elevation data is sufficient for modelling the far away hills. The method followed in this paper (with one panorama taken from the centre of the model) would pose additional manual work to produce a sufficiently accurate panorama. Due to the larger size of the model, and therewith more extreme positions with respect to the panoramic recordings, the angular deviations of objects in the panorama would likely exceed acceptable magnitudes. We plan to provide more accurate maps of the testing grounds in the near future. 

\section*{Acknowledgments}
We thank Thierry Meurers for assisting in the drone-based mapping of the field. This work has, in part, been funded by the Dr.-Klaus-Tschira-Foundation, through Grant No. 00.300.2016 (Robotik  in der Biologie:  Ziele  finden  mit  einem  winzigen  Gehirn. Die neuronalen Grundlagen der Navigation der Bienen).

\nolinenumbers
\clearpage

\begin{appendices}

\setlength\LTleft{-4.5cm}
\setlength\LTright{0pt plus 1fill minus 1fill}
\begin{longtable}{|p{0.6cm}|>{\raggedright}p{2cm}|>{\raggedright}p{2cm}|>{\raggedright}p{2cm}|>{\raggedright}p{1cm}|>{\raggedright}p{1.2cm}|>{\raggedright}p{1.5cm}|p{0.7cm}|p{1.5cm}|}
\hline
\textbf{cit.} & \textbf{ommatidia distribution} & \textbf{ommatidia model} & \textbf{imaging technique} & \textbf{FoV} & \textbf{spectral properties} & \textbf{world model} & \textbf{open} & \textbf{remarks} \\ 

\hline
\cite{giger_honeybee_1996}  & $N_o = 6011$\\ $\Delta \varphi_h = (\SI{2.8}{\degree},\SI{3.7}{\degree})$ , $\Delta \varphi_v = (\SI{1.5}{\degree},\SI{3.5}{\degree})$\\ based on \cite{seidl_visual_1981} and \cite{laughlin_angular_1971} & $\Delta \rho = \SI{2.6}{\degree}$ \\ $N_s = 441$ \\  Gaussian weighting & texture projection on sphere  & $\SI{180}{\degree}\times \SI{180}{\degree}$ & no & 2-D greyscale images  & no & online demo available \footnote{\url{http://andygiger.com/science/beye/beyehome.html}}  
\\ 
\hline
\cite{collins_reconstructing_1997} & $N_o = 4752$ (configurable) 
\\ $\Delta \varphi$ based on 3-D reconstruction of bee eye & $\Delta \rho = 0.82 \degree$ 
\\ $N_s = 8$, Gaussian weighting & raytracing & full FoV of bee & full spectrum & 3-D world & no & 
\\ 
\hline
\cite{neumann_modeling_2002} & $N_o = 2562$, \\$\Delta \varphi$ = \SI{4.3}{\degree}& $\Delta \rho = \SI{5}{\degree}$, Gaussian weighting & remapping, look-up table  & $\SI{360}{\degree}\times \SI{180}{\degree}$ & no & cubic environment map (72x72x6 pixels) & no & \\ 
\hline
\cite{ikeno_reconstruction_2003} & $N_o = 5000$ \\ $\Delta \varphi$ =  \SI{0.5}{\degree}  & $\Delta \rho = \SI{3.54}{\degree}$ \\ Gaussian weighting & luminance based on viewing angle and distance & full FoV of bee and Boundaries based on \cite{seidl_visual_1981} & no & 2-D images & no & \\ 
\hline 
\cite{dickson_integrative_2006} & $N_o = 642$ \\ $\Delta \varphi = (\SI{6.8}{\degree},\SI{9.3}{\degree})$ & $\Delta \rho = \SI{7.48}{\degree}$ \\ Gaussian weighting & same as \cite{neumann_modeling_2002} & n/a  & no & simply structured, textured tunnel & yes & temporal dynamics of photoreceptor \\ 
\hline
\cite{sturzl_mimicking_2010} & Same as \cite{giger_honeybee_1996}, extended to full FoV using \cite{seidl_visual_1981} \\ $\Delta \varphi_h = (2.4,4.6)$, \\ $\Delta \varphi_v = (1.5 , 4.5)$ & $\Delta \rho = (2.6 , 4.5)$, depends on $\Delta \varphi$ \\ $9\times9$ sampling grid & remapping of panoramas & full FoV of bee, boundaries based on \cite{seidl_visual_1981} & no & \SI{360}{\degree} panoramas, or hardware camera (sphere) & no & hardware implementation available 
\\
\hline
\cite{cope_model_2016} & Based on \cite{giger_honeybee_1996} & single ray per ommatidium & ray casting & $\SI{180}{\degree}\times \SI{180}{\degree}$ & no & 3-D shape primitives & yes  & SpineML interface \\ 
\hline
\cite{rodriguezgirones_tobeeview:_2016} & uniform, hexagonal grid & configurable, determines area of image to be sampled & subsampling of input image & n/a   & yes, given weight matrices & 2-D images & yes& configur-able to match focal species \\ 
\hline
\caption{Overview of the properties of different bee eye models. \textit{Ommatidia distribution} describes how the ommatidia are aranged. Parameters are interommatidial angles ($\Delta \varphi$) and number of ommatidia ($N_o$). \textit{Ommatidia model} describes how the individual ommatidia are modeled. Parameters are the number of sampling points per ommatidium ($N_s$) and the acceptance angle ($\Delta \rho$). Most models use a gaussian weighting when integrating samples. \textit{Imaging technique} may vary between remapping of a panorama or cubic environment map, ray tracing, ray casting or via hardware. \textit{FoV} denotes the field of view of the model. \textit{Spectral properties} states whether and how the system models the perception of different wavelengths. In the column \textit{world model} we describe which type of environment can be imaged. \textit{Open source} states if the model and its source are  available for free.} 
\label{table:overview_models}
\end{longtable}

\begin{figure}
\caption{This figure compares the shape of different sampling functions viewed from the side and the top. The top view shows the angular deviation from the main optical axis of the ommatidia ($x$ and $y$), the side view shows the horizontal deviation ($x$) and the corresponding weights ($w$), the black lines depict the FWHM. The weights determine how much the sampled color contributes to the perceived color of the ommatidium. The Stürzl-model uses a square sampling function with 81 sampling points, and Giger uses a concentric disk sampling function with 441 sampling points. How the number of samples affect the output image is thoroughly compared in the results chapter.}
\includegraphics[width=\textwidth]{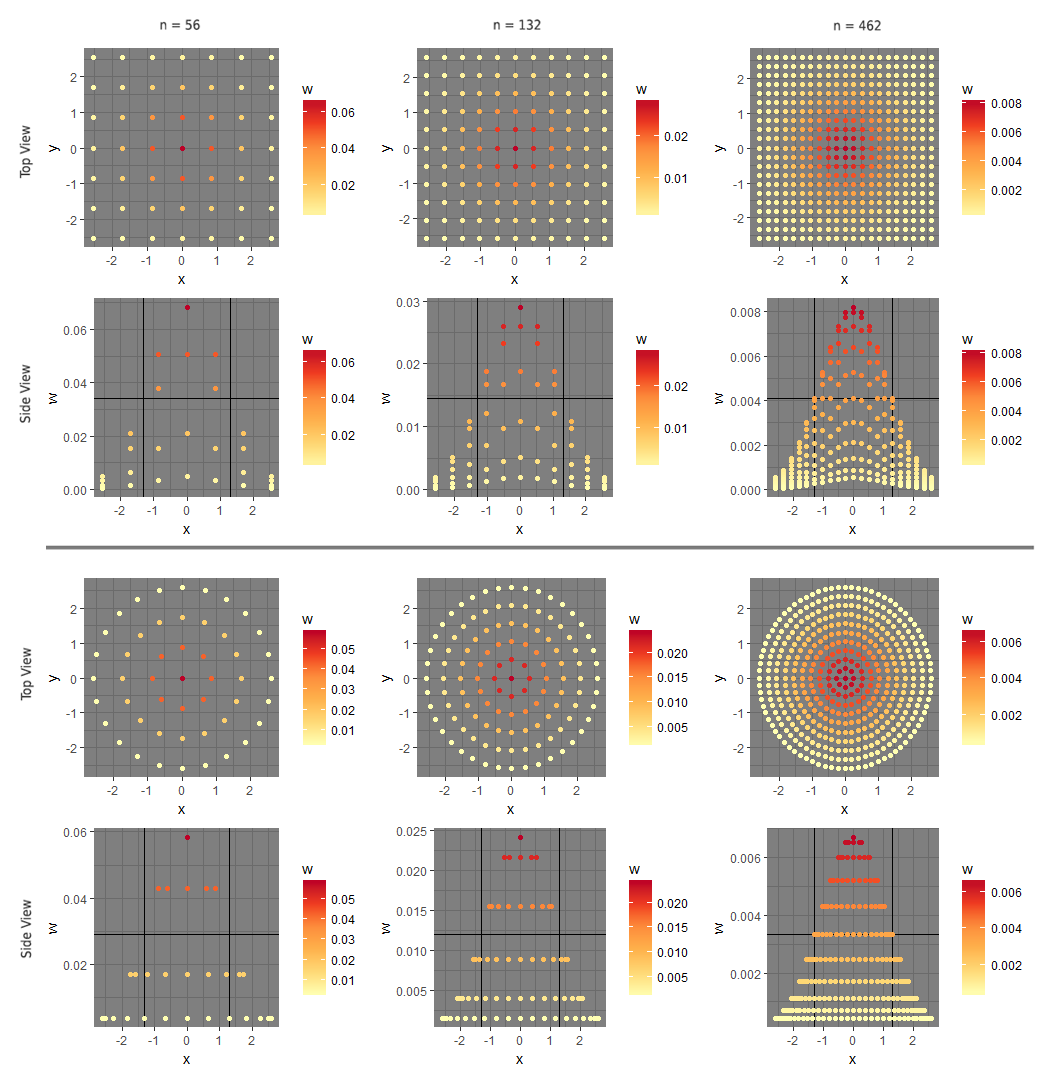}
\centering
\label{fig:comparison_acceptance_functions}
\end{figure}

\end{appendices}
\clearpage

\bibliography{main}

\begin{thebibliography}{10}

\bibitem{anderson_shape_1977}
A.~M. Anderson.
\newblock Shape perception in the honey bee.
\newblock {\em Animal Behaviour}, 25:67--79, Feb. 1977.

\bibitem{benthin_efficient_2015}
C.~Benthin, S.~Woop, M.~Nießner, K.~Selgrad, and I.~Wald.
\newblock Efficient {Ray} {Tracing} of {Subdivision} {Surfaces} using
  {Tessellation} {Caching}.
\newblock In {\em Proceedings of the 7th {High}-{Performance} {Graphics}
  {Conference}}. ACM, 2015.

\bibitem{borst_drosophilas_2009}
A.~Borst.
\newblock Drosophila's {View} on {Insect} {Vision}.
\newblock {\em Current Biology}, 19(1):R36--R47, Jan. 2009.

\bibitem{cheeseman2014way}
J.~F. Cheeseman, C.~D. Millar, U.~Greggers, K.~Lehmann, M.~D. Pawley, C.~R.
  Gallistel, G.~R. Warman, and R.~Menzel.
\newblock Way-finding in displaced clock-shifted bees proves bees use a
  cognitive map.
\newblock {\em Proceedings of the National Academy of Sciences}, page
  201408039, 2014.

\bibitem{cheung2014still}
A.~Cheung, M.~Collett, T.~S. Collett, A.~Dewar, F.~Dyer, P.~Graham, M.~Mangan,
  A.~Narendra, A.~Philippides, W.~St{\"u}rzl, et~al.
\newblock Still no convincing evidence for cognitive map use by honeybees.
\newblock {\em Proceedings of the National Academy of Sciences},
  111(42):E4396--E4397, 2014.

\bibitem{collett2002memory}
T.~S. Collett and M.~Collett.
\newblock Memory use in insect visual navigation.
\newblock {\em Nature Reviews Neuroscience}, 3(7):542, 2002.

\bibitem{collins_reconstructing_1997}
S.~Collins.
\newblock Reconstructing the {Visual} {Field} of {Compound} {Eyes}.
\newblock In {\em Rendering {Techniques} ’97}, Eurographics, pages 81--92.
  Springer, Vienna, 1997.

\bibitem{cope_model_2016}
A.~J. Cope, C.~Sabo, K.~Gurney, E.~Vasilaki, and J.~A.~R. Marshall.
\newblock A {Model} for an {Angular} {Velocity}-{Tuned} {Motion} {Detector}
  {Accounting} for {Deviations} in the {Corridor}-{Centering} {Response} of the
  {Bee}.
\newblock {\em PLOS Computational Biology}, 12(5):e1004887, May 2016.

\bibitem{cruse2011no}
H.~Cruse and R.~Wehner.
\newblock No need for a cognitive map: decentralized memory for insect
  navigation.
\newblock {\em PLoS computational biology}, 7(3):e1002009, 2011.

\bibitem{dickson_integrative_2006}
W.~Dickson, A.~Straw, C.~Poelma, and M.~Dickinson.
\newblock An {Integrative} {Model} of {Insect} {Flight} {Control} ({Invited}).
\newblock In {\em 44th {AIAA} {Aerospace} {Sciences} {Meeting} and {Exhibit}},
  Reno, Nevada, Jan. 2006. American Institute of Aeronautics and Astronautics.

\bibitem{floreano_miniature_2013}
D.~Floreano, R.~Pericet-Camara, S.~Viollet, F.~Ruffier, A.~Brückner,
  R.~Leitel, W.~Buss, M.~Menouni, F.~Expert, R.~Juston, M.~K. Dobrzynski,
  G.~L’Eplattenier, F.~Recktenwald, H.~A. Mallot, and N.~Franceschini.
\newblock Miniature curved artificial compound eyes.
\newblock {\em Proceedings of the National Academy of Sciences},
  110(23):9267--9272, June 2013.

\bibitem{giger_honeybee_1996}
A.~Giger.
\newblock {\em Honeybee vision: analysis of pattern orientation}.
\newblock Doctoral {Thesis}, Australian National University, Canberra,
  Australia, 1996.

\bibitem{eigen_community_eigen_nodate}
G.~Guennebaud, B.~Jacob, et~al.
\newblock Eigen v3.
\newblock http://eigen.tuxfamily.org, 2010.

\bibitem{ikeno_reconstruction_2003}
H.~Ikeno.
\newblock A reconstruction method of projection image on worker honeybees’
  compound eye.
\newblock {\em Neurocomputing}, 52-54:561--566, June 2003.

\bibitem{laughlin_angular_1971}
S.~B. Laughlin and G.~A. Horridge.
\newblock Angular sensitivity of the retinula cells of dark-adapted worker bee.
\newblock {\em Zeitschrift für Vergleichende Physiologie}, 74(3):329--335,
  1971.

\bibitem{menzel2011common}
R.~Menzel, A.~Kirbach, W.-D. Haass, B.~Fischer, J.~Fuchs, M.~Koblofsky,
  K.~Lehmann, L.~Reiter, H.~Meyer, H.~Nguyen, et~al.
\newblock A common frame of reference for learned and communicated vectors in
  honeybee navigation.
\newblock {\em Current Biology}, 21(8):645--650, 2011.

\bibitem{muller_familiarity_2016}
J.~Müller.
\newblock Familiarity in {Honeybee} {Navigation} - {Behavioral} and
  {Neurocomputational} {Investigations}.
\newblock Master's thesis, Ludwig-Maximilians-Universität München and Freie
  Universität Berlin, Munich, 2016.

\bibitem{neumann_modeling_2002}
T.~R. Neumann.
\newblock Modeling {Insect} {Compound} {Eyes}: {Space}-{Variant} {Spherical}
  {Vision}.
\newblock In {\em Biologically {Motivated} {Computer} {Vision}}, Lecture
  {Notes} in {Computer} {Science}, pages 360--367. Springer, Berlin,
  Heidelberg, Nov. 2002.

\bibitem{riley1996tracking}
J.~Riley, A.~Smith, D.~Reynolds, A.~Edwards, J.~Osborne, I.~Williams,
  N.~Carreck, and G.~Poppy.
\newblock Tracking bees with harmonic radar.
\newblock {\em Nature}, 379(6560):29, 1996.

\bibitem{rodriguezgirones_tobeeview:_2016}
M.~A. Rodríguez‐Gironés and A.~Ruiz.
\newblock {toBeeView}: a program for simulating the retinal image of visual
  scenes on nonhuman eyes.
\newblock {\em Ecology and Evolution}, 6(21):7892--7900, 2016.

\bibitem{rosca_new_2010}
D.~Roşca.
\newblock New uniform grids on the sphere.
\newblock {\em Astronomy and Astrophysics}, 520:A63, Sept. 2010.

\bibitem{seidl_visual_1981}
R.~Seidl and W.~Kaiser.
\newblock Visual field size, binocular domain and the ommatidial array of the
  compound eyes in worker honey bees.
\newblock {\em Journal of comparative physiology}, 143(1):17--26, Mar. 1981.

\bibitem{srinivasan_pattern_1994}
M.~V. Srinivasan.
\newblock Pattern recognition in the honeybee: {Recent} progress.
\newblock {\em Journal of Insect Physiology}, 40(3):183--194, Mar. 1994.

\bibitem{srinivasan2014going}
M.~V. Srinivasan.
\newblock Going with the flow: a brief history of the study of the honeybee’s
  navigational ‘odometer’.
\newblock {\em Journal of Comparative Physiology A}, 200(6):563--573, 2014.

\bibitem{sturzl_mimicking_2010}
W.~Stürzl, N.~Boeddeker, L.~Dittmar, and M.~Egelhaaf.
\newblock Mimicking honeybee eyes with a 280 degrees field of view catadioptric
  imaging system.
\newblock {\em Bioinspiration \& Biomimetics}, 5(3):036002, Sept. 2010.

\bibitem{sturzl_three-dimensional_2015}
W.~Stürzl, I.~Grixa, E.~Mair, A.~Narendra, and J.~Zeil.
\newblock Three-dimensional models of natural environments and the mapping of
  navigational information.
\newblock {\em Journal of Comparative Physiology A}, 201(6):563--584, June
  2015.

\bibitem{varela_optics_1970}
F.~G. Varela and W.~Wiitanen.
\newblock The {Optics} of the {Compound} {Eye} of the {Honeybee} ({Apis}
  mellifera).
\newblock {\em The Journal of General Physiology}, 55(3):336--358, Mar. 1970.

\bibitem{von1967dance}
K.~Von~Frisch.
\newblock The dance language and orientation of bees.
\newblock 1967.

\bibitem{afra_embree_2016}
A.~T. Áfra, I.~Wald, C.~Benthin, and S.~Woop.
\newblock Embree ray tracing kernels: overview and new features.
\newblock pages 1--2. ACM Press, 2016.

\end{thebibliography}

\bibliographystyle{abbrv}

\end{document}